\documentclass[11pt]{article} 

\usepackage{color,graphicx}

\usepackage{young}
\usepackage[vcentermath]{youngtab}

\usepackage{amsmath,amssymb,graphicx}
\usepackage{hyperref}

\definecolor{darkred}{rgb}{0.65,0.15,0}
\hypersetup{pdfborder={0 0 0},colorlinks=true,urlcolor=darkred,citecolor=blue,linkcolor=darkred,linktocpage=true}

\newcommand{\eprint}[1]{{\href{http://arxiv.org/abs/#1}{\texttt{[#1}]}}}
\newcommand{\eprintN}[1]{{\href{http://arxiv.org/abs/#1}{\texttt{#1 [hep-th]}}}}
\newcommand{\doi}[2]{{\hypersetup{urlcolor=darkred}\href{http://dx.doi.org/#2}{#1}\hypersetup{urlcolor=blue}}}

\usepackage{cite}

\newcommand{\nn}{\nonumber}
\newcommand{\mf}[1]{\mathfrak{#1}}

\newcommand{\Minfty}{\textrm{Maxwell}_{\infty}}
\newcommand{\MA}{\textrm{Maxwell}}

\newcommand{\lb}{\left[}
\newcommand{\rb}{\right]}

\newcommand{\height}{\mathrm{ht}}
\newcommand{\mult}{\mathrm{mult}}
\newcommand{\chr}{\mathrm{dom\_char}}

\newcommand{\cone}{c_1}
\newcommand{\ctwo}{c_2}
\newcommand{\cn}{c_n}

\makeatletter

\@addtoreset{equation}{section}
\makeatother

\begin{document}

\thispagestyle{empty}

\hfill{ICCUB-17--002}

\vspace{10mm}

\begin{center}
{\LARGE \bf On free Lie algebras and particles\\[5mm] in electro-magnetic fields}\\[10mm]

\vspace{8mm}
\normalsize
{\large  Joaquim Gomis${}^{1}$ and Axel Kleinschmidt${}^{2,3}$\\[2mm]}

\vspace{10mm}
${}^1${\it Departament de F\'isica Qu\`antica i Astrof\'isica and \\Institut de Ci\`encies del Cosmos (ICCUB), 
Universitat de Barcelona, \\Mart\'i i Franqu\`es 1, E-08028 Barcelona, Spain}
\vskip1em
${}^2${\it Max-Planck-Institut f\"{u}r Gravitationsphysik (Albert-Einstein-Institut)\\
Am M\"{u}hlenberg 1, DE-14476 Potsdam, Germany}
\vskip 1 em
${}^3${\it International Solvay Institutes\\
ULB-Campus Plaine CP231, BE-1050 Brussels, Belgium}
\vspace{15mm}

\hrule

\vspace{10mm}

\begin{tabular}{p{12cm}}
{\small
The Poincar\'e algebra can be extended (non-centrally) to the Maxwell algebra and beyond. These extensions are relevant for describing particle dynamics in electro-magnetic backgrounds and possibly including the backreaction due the presence of multipoles. We point out a relation of this construction to free Lie algebras that gives a unified description of all possible kinematic extensions, leading to a symmetry algebra that we call $\Minfty$. A specific dynamical system with this infinite symmetry is constructed and analysed.
}
\end{tabular}
\vspace{10mm}
\hrule
\end{center}

\newpage
\setcounter{page}{1}

\tableofcontents
 
 \vspace{5mm}
 \hrule
 \vspace{5mm}
 
\section{Introduction}
The motion of a system of electromagnetic charges in special relativity in terms of multipoles is a well-known subject, see for example \cite{Dixon}. The motion could be described in terms of the world line of the center of mass and the multipoles of the distribution of charges.\footnote{For a single (monopole moment) charge, the problem was also studied in~\cite{Taub:1948zza}.}

The space-time symmetries of particles moving in an electro-magnetic background is much less known.  To our knowledge the first attempt to study these symmetries was in the framework of kinematical algebras \cite{Bacry:1968zf} and was done by Bacry--Combe--Richard (BCR) in \cite{Bacry:1970ye} for the case of a particle in a given fixed constant electromagnetic field  $F_{ab}$.  The Lorentz generators that leave $F_{ab}$ invariant are two combinations of Lorentz generators
\begin{equation}
\label{bcr}
 G=\frac 12  F^{ab} M_{ab} \,,\quad\quad
{}\star G=\frac 12 {}\star F^{ab} M_{ab}\,.
\end{equation}
The BCR algebra has 6 generators, four translations and two of the previous Lorentz transformations. 
The algebra has two central charges
\begin{equation}
\label{bcr1}
 \lb P_a, P_b \rb=Z F^{ab} M_{ab} +Z{}\star F^{ab} M_{ab}
\end{equation}
and the two central charges are identified with electric and magnetic charge.
Later, the Maxwell algebra was introduced  by Schrader \cite{Schrader:1972zd}, where the electro-magnetic field $F_{ab}$ was allowed to transform with the Lorentz and the Poincar\'e algebra and then received six non-central extensions\footnote{It is well-known that the $D$-dimensional Poincar\'e algebra admits non-central extensions~\cite{Galindo}.}
\begin{align}
\label{M0}
\lb P_a, P_b \rb = Z_{ab}\,.
\end{align}
This Maxwell algebra appears in~\cite{Bonanos:2008ez} as the symmetry group of a particle moving in a generic constant electro-magnetic background. The Maxwell algebra describes at same time the particle and the constant electro-magnetic background where the particle moves. In~\cite{Bonanos:2008ez} an infinite extension of the Maxwell algebra ($\Minfty$) was also envisioned through the process of extending by non-trivial two-forms in the Chevalley--Eilenberg Lie algebra cohomology. This was done in an iterative procedure and the full mathematical structure of $\Minfty$ was not uncovered.

In this paper, we find the mathematical structure underlying $\Minfty$. It turns out to be the semi-direct product of the Lorentz algebra with a free Lie algebra generated by the Poincar\'e translations and the algebra is $\mathbb{Z}$-graded (with empty negative levels). The finite level extensions found  in~\cite{Bonanos:2008ez} are easily reproduced and the associated Young tableaux are identified. This construction works in any space-time dimension. We present the lowest level generators of the algebra in table~\ref{tab:intro} for the four-dimensional case. From the free algebra one can in principle compute any finite level extension and the corresponding commutation relations. We also study possible quotients of $\Minfty$. As is clear from table~\ref{tab:intro} and~\eqref{M0}, the Maxwell algebra studied by Schrader~\cite{Schrader:1972zd} corresponds to the quotient of $\Minfty$ where one only keeps levels $\ell=0$, $\ell=1$ and $\ell=2$.

\renewcommand{\arraystretch}{1.5}
\begin{table}[t!]
\centering
\caption{\it Summary of the low level generators of $\Minfty$ in $D=4$ dimensions. The Young tableau describes the tensor type as a representation of the symmetric group. All generators are Lorentz tensors under the anti-symmetric $M_{ab}$. Levels $\ell=0$ and $\ell=1$ together constitute the Poincar\'e algebra and the translation generators $P_a$ generate the higher levels freely (in a Lie algebraic sense). We also display the coordinates that are associate with the generators in the non-linear realisation of $\Minfty/$Lorentz.\label{tab:intro}}
\begin{tabular}{c||c|c|c}
Level & Young tableau & Generator and some commutators & coordinates\\
\hline\hline
$\ell=0$ & \raisebox{0.2\height}{\scalebox{0.6}{$\yng(1,1)$}} & $M_{ab}$ & none\\\hline
$\ell=1$ & \raisebox{0.2\height}{\scalebox{0.6}{$\yng(1)$}} & $P_a$ & $x^a$ \\\hline
$\ell=2$ & \raisebox{0.2\height}{\scalebox{0.6}{$\yng(1,1)$}} & $Z_{ab}=\lb P_a,P_b \rb$ & $\theta^{ab}$ \\\hline
$\ell=3$ &  \raisebox{0.2\height}{\scalebox{0.6}{$\yng(2,1)$}}  & $Y_{ab,c}=\lb Z_{ab}, P_c\rb$ & $\xi^{ab,c}$ \\\hline
$\ell=4$ & \raisebox{0.1\height}{\scalebox{0.6}{$\yng(3,1)$}} & $S_{ab,c,d}^1$, see~\eqref{eq:inv4} & $\sigma_1^{ab,c,d}$ \\[3mm]
&  \raisebox{-0.1\height}{\scalebox{0.6}{$\yng(2,1,1)$}}  & $S_{abc,d}^2$, see~\eqref{eq:inv4} & $\sigma_2^{abc,d}$ 
\end{tabular}
\end{table}

We construct a dynamical system with $\Minfty$ symmetry. The model that we analyze is lowest order in derivatives and could therefore be considered as the first term of an effective description of an electro-magnetic interaction of particles. Possible higher derivative extensions are not treated in this article. The basic tool that we use is the non-linear realisation in terms of the coset $\Minfty/\mathrm{Lorentz}$ and the associated Maurer--Cartan forms. The coset can described in terms of the generalized coordinates  $x^a, \theta^{ab}, \xi^{ab,c},  \sigma_1^{ab,c,d}, \ldots$ dual to the generators at levels $\ell>0$. Mathematically, these coordinates are just a local parametrisation of the infinite-dimensional coset in a Lorentz
gauge-fixed form. We will also assign physical significance to them by thinking of $x^a$ as giving the space-time coordinate (of the center of mass) of a charge distribution and the additional coordinates could be considered either as describing higher inertial multipoles of a charge distribution or as coordinates on some generalized space-time. This latter viewpoint has been taken repeatedly~\cite{Sohnius:1978fw,Duff:1989tf,Gauntlett:1990nk,Duff:1990hn,deWit:2000wu} and is particularly pronounced in recent work on exceptional symmetries in supergravity~\cite{West:2003fc,Kleinschmidt:2003jf,Hull:2007zu,Coimbra:2011ky,Coimbra:2012af,Berman:2012vc,Hohm:2013uia}.
We also require the introduction of an infinite set new dynamical variables 
$f_{ab}, f_{ab,c}, f_{ab,c,d}^1, \cdots$ that make it possible to write a manifest Lorentz invariant Lagrangian for the system. These quantities are also the canonical momenta associated to the generalized coordinates. By formal similarity of our equations to those of~\cite{Dixon}, the extra dynamical variables are related to the higher electro-magnetic multipole moments of the system of charges while the coordinates $\theta^{ab}$, $\xi^{ab,c}$,\ldots are similar to higher inertial moments like angular momentum. However, this assignment is based only on similarity and we consider further analysis of the precise interpretation of the higher coordinates and dynamical variables to be necessary. 

The dynamical equations of motion of our system relate the extra dynamical variables with the coordinates on $\Minfty/\mathrm{Lorentz}$. A universal equation motion that always is present in our dynamical system is\footnote{We have used the proper time of the center of mass as the evolution parameter.}
\begin{align}
m\ddot{x}_a = f_{ab} \dot{x}^b\,.
\end{align}
This equation is the Lorentz force for our system. Note that our differential equations of motion are  $\Minfty$ invariant. When we consider a particular solution of the equations of motion for $f_{ab}, f_{ab,c}, f_{ab,c,d}^1, \ldots$, we can see that $f_{ab}$ is now a function in general of the generalized coordinates $f_{ab}\rightarrow  F_{ab}(x,\theta,\ldots)$ and the symmetry $\Minfty$ is spontaneously broken.

It is not clear whether $F_{ab}$ can always be interpreted as an electro-magnetic field in ordinary space-time that satisfies the integrability (or Bianchi) identity $\partial_{[a} F_{bc]}=0$.\footnote{The violation of the Bianchi identity in ordinary configuration space could be associated with magnetic monopoles.}  We confirm the finding of~\cite{Bonanos:2008ez} that this is not always the case. The electro-magnetic field $F_{ab}(x,\theta\cdots)$ appearing in the Lorentz equation is not necessarily integrable if one considers all of $\Minfty$. Restricting the free Lie algebra to a suitable quotient renders the field integrable and in fact the quotient just corresponds to the unfolding formalism for the Maxwell field studied for example in~\cite{Vasiliev:2005zu,Boulanger:2015mka}. We can thus describe the motion of a charged particle in a generic (analytic) background. Our formulation gives a different view on the standard electro-dynamics of a point particle and potentially also describes generalizations for systems of charges in terms of inertial moments.

Going beyond the quotient related to the unfolding formalism one encounters additional dynamical equations and non-integrable solutions to the Maxwell equation. These appear to correspond to the back-reaction of the multipoles on the motion of the center of mass.

The organisation of the paper is as follows: In section~\ref{sec:MA}, we review the non-central extension process for the Poincar\'e algebra. Section~\ref{sec:FLA} contains the description of free Lie algebras and the isomorphism of the free Lie algebra generated by the translations $P_a$ to the extension obtained in section~\ref{sec:MA}. We also discuss quotients of $\Minfty$. In section~\ref{sec:dyn}, we construct a dynamical point particle model with $\Minfty$ symmetry and analyse its equations of motion. The relation to electrodynamics and unfolded dynamics is discussed. We offer some concluding remarks in section~\ref{sec:concl} and collect some background material on free Lie algebras in an appendix.

\section{Extensions of the Poincar\'e algebra}
\label{sec:MA}

In this section, we briefly review the extension of the Poincar\'e algebra based on Eilenberg--Chevalley cohomology. The results of this section were obtained in~\cite{Bonanos:2008ez}.

The Poincar\'e algebra in $D$ space-time dimensions is a semi-direct sum of the Lorentz algebra $\mf{so}(1,D-1)$ and the abelian algebra of space-time translations. We denote the Lorentz generators by $M_{ab}=M_{[ab]}$ for $a=0,\ldots, D-1$ and the translation generators by $P_a$. Their Poincar\'e Lie algebra is
\begin{align}
\label{eq:PA}
\lb M_{ab} , M_{cd} \rb &= \eta_{bc} M_{ad} -\eta_{bd} M_{ac} -\eta_{ac} M_{bd} + \eta_{ad} M_{bc}\,,\nn\\
\lb M_{ab}, P_c \rb &= \eta_{bc} P_a - \eta_{ac} P_b\,,\nn\\
\lb P_a, P_b \rb &=0\,.
\end{align}
Here, $\eta_{ab}=(-+\ldots+)$ is the flat Minkowski metric. We will refer to the Lorentz generators as level $\ell=0$ and the Poincar\'e generators as level $\ell=1$.\footnote{This terminology differs from the one in~\cite{Bonanos:2008ez} but it turns out to be more convenient for the connection to free Lie algebras discussed in the present paper.}

In order to determine possible extensions of this Lie algebra one can study the Chevalley--Eilenberg cohomology~\cite{AzcarragaIzquierdo}. It turns out~\cite{Bonanos:2008ez} that there is a sequence of extensions of the algebra by generators that are Lorentz tensors and in fact can be viewed also as tensors of the general linear algebra $\mf{gl}(D)$ and represented by Young tableaux. In this representation, the translation generators $P_a$ of level $\ell=1$ are written as a single box
\begin{align}
\textrm{Level $\ell=1$:}\quad\quad P_a \longleftrightarrow \yng(1)\,.
\end{align}
We will also assign level $\ell=0$ to the Lorentz generators $M_{ab}$.

The Poincar\'e algebra has non-trivial cohomology that can be parametrised by the anti-symmetric tensor $Z_{ab}=Z_{[ab]}$~\cite{Schrader:1972zd,Bonanos:2008ez}.\footnote{Here and everywhere in the paper we use (anti-)symmetrizations of strength one.} Adding this generator to the Poincar\'e algebra deforms the commutation relations in~\eqref{eq:PA} to
\begin{align}
\label{eq:M2}
\lb P_a, P_b \rb = Z_{ab}\,,\quad\quad
\lb Z_{ab}, Z_{cd} \rb =0\,,\quad\quad
\lb Z_{ab}, P_c \rb =0\,.
\end{align}
The tensorial nature of $Z_{ab}$ is expressed by the commutation relation
\begin{align}
\lb M_{ab} , Z_{cd} \rb &= \eta_{bc} Z_{ad} -\eta_{bd} Z_{ac} -\eta_{ac} Z_{bd} + \eta_{ad} Z_{bc}
\end{align}
with the Lorentz generators $M_{ab}$. We will refer to the generator $Z_{ab}$ as level $\ell=2$ and as a Young tableaux it is given by
\begin{align}
\textrm{Level $\ell=2$:}\quad\quad Z_{ab} \longleftrightarrow \yng(1,1)\,.
\end{align}
The algebra generated by $(M_{ab}, P_a, Z_{ab})$ closes and is commonly called the \textit{Maxwell algebra}~\cite{Schrader:1972zd}. In view of the sequence of further extensions of the Poincar\'e algebra we will refer to it as $\MA_2$ as it used the generators up to level $\ell=2$. The connection between this algebra and particle motion in a constant electro-magnetic background $F_{ab}=\textrm{const.}$ has been well-studied~\cite{Bonanos:2008ez}, see also~\cite{Schrader:1972zd,Bacry:1970ye,Bacry:1970du}. We note that the commutators~\eqref{eq:M2} are consistent with the level grading that we have assigned to the generators. The commutator $\lb P_a, P_b\rb =Z_{ab}$ has two generators of level $\ell=1$ on the left-hand side and the right-hand side a single generator of level $\ell=2$. The vanishing $\lb Z_{ab}, Z_{cd}\rb=\lb Z_{ab},P_c\rb=0$ within the algebra $\MA_2$ then is simply due to the fact that there are no generators of level $\ell>2$ in $\MA_2$.

The cohomological analysis can be repeated and one finds that there is also a non-trivial cohomology of the algebra $\MA_2$~\cite{Bonanos:2008ez}. One can therefore extend $\MA_2$ to an algebra $\MA_3$ by introducing new generators 
\begin{align}
\textrm{Level $\ell=3$:}\quad\quad Y_{ab,c} \longleftrightarrow \yng(2,1)
\end{align}
that we will call generators of level $\ell=3$ (since their Young tableaux has three boxes). They can appear in the commutation relations consistently with the level grading and we will see below how they change some of the commutators in~\eqref{eq:M2}.

The Young tableaux above has mixed symmetry and since we will encounter many such Young tableaux we will now fix our conventions for labelling tensor generators associated with them: We transverse the Young tableau by column from left to right; each column corresponds to an anti-symmetric set of indices. We separate the columns (=sets of anti-symmetric indices) by commas. A tensor associated with such a Young tableau is therefore automatically anti-symmetric in every set of indices. The irreducibility (as a $\mf{gl}(D)$ representation) of the representation encoded in the Young tableaux is equivalent to the requirement that the anti-symmetrization of the indices of one column with any single index of a column to the right gives zero. If there are columns of equal length, the tensor is symmetric under interchange of the sets of indices. A discussion of Young symmetrizers and representations of the symmetric group based on tableaux can be found for example in~\cite{Fulton}.

In the example above this means that $Y_{ab,c}$ has the following symmetry properties
\begin{align}
Y_{[ab],c} = Y_{ab,c}\,,\quad Y_{[ab,c]} =0\,.
\end{align}
It arises in the commutator\footnote{The order of indices on $Y_{ab,c}$ and the normalisation of the generator differs (by $3$) from the one used in~\cite{Bonanos:2008ez}.}
\begin{align}
\label{eq:L3MA}
\lb Z_{ab} , P_c \rb = Y_{ab,c}\,.
\end{align}
This commutator automatically satisfies that the totally anti-symmetric part vanishes by the Jacobi identity upon substitution of~\eqref{eq:M2}. Moreover, $Y_{ab,c}$ transforms as a tensor under the Lorentz generators $M_{ab}$ in the standard way
\begin{align}
\lb M_{ab} , Y_{cd,e} \rb = \eta_{bc} Y_{ad,e} - \eta_{ac} Y_{bd,e} + \eta_{bd} Y_{ca,e} -\eta_{ad} Y_{cb,e} + \eta_{be} Y_{cd,a}-\eta_{ae} Y_{cd,b}\,.
\end{align}
We will denote the Lie algebra generated by $(M_{ab}, P_a, Z_{ab}, Y_{ab,c})$ by $\MA_3$. The generator $Y_{ab,c}$ of level $\ell=3$ commutes with all generators of levels $\ell\ge 1$ if one considers $\MA_3$. This is again due to the fact that our level assignment provides a consistent grading of the Lie algebra $\MA_3$.

Performing the cohomological analysis of $\MA_3$ one finds again that it admits an extension~\cite{Bonanos:2008ez}.\footnote{We assume $D\geq 4$ here for simplicity.} This time the extending generators belong to two different irreducible representations of $\mf{gl}(D)$. They are given by\footnote{Also here the convention for the order of indices differs from~\cite{Bonanos:2008ez}. The reason is that we want to maintain our general labelling convention for Young tableaux.}
\begin{align}
\label{eq:M4}
\textrm{Level $\ell=4$:}\quad\quad S_{ab,c,d}^1 \longleftrightarrow \yng(3,1)\quad\textrm{and}\quad S_{abc,d}^2 \longleftrightarrow \yng(2,1,1)\,.
\end{align}
Even though the generators could be distinguished by their tensor structure, we have put superscripts on them to make them easier to distinguish. In agreement with our rules above, the generators have the following tensor properties
\begin{align}
S^1_{[ab],c,d} = S^1_{ab,c,d}\,,\quad
S^1_{ab,(c,d)} = S^1_{ab,c,d}\,,\quad
S^1_{[ab,c],d} = 0
\end{align}
and
\begin{align}
S^2_{[abc],d} = S^2_{abc,d}\,,\quad
S^2_{[abc,d]} = 0\,.
\end{align}
The new generators arise in the commutators as follows:
\begin{align}
\label{eq:L4MA}
\lb Y_{ab,c},P_d \rb &= S^1_{ab,c,d}  + 2S^2_{abd,c}- S^2_{bcd,a}- S^2_{cad,b}\nn\\
&= S^1_{ab,c,d} +3 S^2_{dab,c} - S^2_{abc,d}\,,
\end{align}
where we have written the right-hand side in two different ways using the irreducibility constraint $S^2_{[abc,d]}=0$ of the second generator arising at level $\ell=4$. This relation fixes also 
\begin{align}
\lb Z_{ab} , Z_{cd} \rb = 
 3S^2_{abd,c} -S^2_{abc,d} - 3 S^2_{abc,d}+S^2_{abd,c} 
= -8 S^2_{ab[c,d]}
\end{align}
by Jacobi identities. All remaining commutators are also completely determined by the level grading and the tensor structure of the generators. The Lie algebra that is generated by $(M_{ab}, P_a, Z_{ab}, Y_{ab,c}, S_{ab,c,d}^1, S_{abc,d}^1)$ will be called $\MA_4$. For completeness, we also note the inverse relations
\begin{align}
\label{eq:inv4}
S^2_{abc,d} &= -\frac38 \lb Z_{[ab}, Z_{c]d}\rb   = -\frac38 \lb P_{[c}, Y_{ab],d}\rb\,,\nn\\
S^1_{ab,c,d} &= \frac38 \lb P_d, Y_{ab,c} \rb + \frac38 \lb P_c, Y_{ab,d} \rb -\frac14 \lb P_{[a} , Y_{b]c,d} \rb -\frac14 \lb P_{[a} , Y_{b]d,c} \rb
\end{align}
The tableaux and their dimensions are summarized for $D=4$ in table~\ref{tab:L4}.

\renewcommand{\arraystretch}{1.5}
\begin{table}[t!]
\centering
\caption{\it Summary of the positive level generators for $\MA_4$ in $D=4$ dimensions.\label{tab:L4}}
\begin{tabular}{c||c|c|c}
Level & Young tableau & Generator & Dimension\\
\hline\hline
$\ell=1$ & \raisebox{0.2\height}{\scalebox{0.6}{$\yng(1)$}} & $P_a$ & $4$\\\hline
$\ell=2$ & \raisebox{0.2\height}{\scalebox{0.6}{$\yng(1,1)$}} & $Z_{ab}$ & $6$\\\hline
$\ell=3$ & \raisebox{0.2\height}{\scalebox{0.6}{$\yng(2,1)$}} & $Y_{ab,c}$ & $20$\\\hline
$\ell=4$ &\raisebox{0.1\height}{\scalebox{0.6}{$\yng(3,1)$}}  & $S_{ab,c,d}^1$ & $45$\\[3mm]
& \raisebox{-0.1\height}{\scalebox{0.6}{$\yng(2,1,1)$}} & $S_{abc,d}^2$ & $15$
\end{tabular}
\end{table}

This cohomological process could now be continued and~\cite{Bonanos:2008ez} gives the tableaux for the generators at level $\ell=5$. Rather than pursuing further the step by step cohomological analysis, we now identify the full Lie algebraic structure in slightly different terms.

\section{Free Lie algebras and their quotients}
\label{sec:FLA}

The sequence of Maxwell algebras $\MA_n$ reviewed in the previous section exhibits an intriguing pattern: The generators at level $0<\ell\le n$ are given by tensors of $\mf{gl}(D)$ that transform according to Young tableaux with $\ell$ boxes. The generators at level $\ell+1$ can be obtained from the ones at level $\ell$ upon commutation with the $\ell=1$ generators $P_a$ and all commutators are consistent the level grading. 

Inspection of the Young tableaux that arise up to $\MA_5$ shows that the structure is fully consistent with quotients of a free Lie algebra on $D$ generators. These $D$ generators are precisely the translation generators $P_a$ for $a=0,\ldots, D-1$. This is one of the central results of this paper. Before proving it we review for completeness some basic features of free Lie algebras. More details can be found in appendix~\ref{app:FLA} and~\cite{BourbakiFree,Viennot,KleinschmidtThesis}.

\subsection{Free Lie algebras}

For any (finite) set of independent generators $P_a$ one can define a free Lie algebra $\mf{f}$ as follows. One considers the linear space spanned by all possible multi-commutators $\lb \lb \lb P_{a_1}, P_{a_2} \rb,\ldots P_{a_{\ell-1}}\rb , P_{a_\ell}\rb$ and identifies all elements that are related to each other by the relations of the Lie bracket, namely anti-symmetry and the Jacobi identity. When we refer to this general $\ell$-fold multiple commutator we will sometimes denote it by $Y_{a_1\ldots a_\ell}$ below and in this case the tensorial symmetry properties of the generator are not specified. Since no relations other than anti-symmetry and the Jacobi identity are used this is the minimal requirement on a Lie algebra and the space spanned by all these multi-commutators is therefore called the free Lie algebra $\mf{f}$ generated by the $P_a$. Since in particular, there are no relations that change the number of the $P_a$ in a multi-commutator one can consider the free Lie algebra $\mf{f}$ for a fixed number $\ell$ of elements in the multi-commutator. This is called the level $\ell$ part $\mf{f}_\ell$ of the free Lie algebra that is a direct sum
\begin{align}
\label{eq:fgrad}
\mf{f} = \bigoplus_{\ell>0} \mf{f}_\ell = \mf{f}_1 \oplus \mf{f}_2 \oplus \ldots\,.
\end{align}
More precisely, $\mf{f}_1$ is the $D$-dimensional vector space spanned by the $P_a$. The space $\mf{f}_2$ is the space spanned by all commutators $\lb P_a, P_b\rb$. Due to the anti-symmetry of the Lie bracket this space is of dimension $D(D-1)/2$. We can write this also as $\mf{f}_2 = \lb \mf{f}_1, \mf{f}_1\rb$. As a vector space, $\mf{f}_2$ is the exterior square of $\mf{f}_1$, but we prefer to use the bracket notation since this extends to arbitrary level\footnote{This statement contains non-trivial information: Not all multi-commutators with $\ell$ elements are of the form $\lb \lb \lb P_{a_1}, P_{a_2} \rb,\ldots P_{a_{\ell-1}}\rb , P_{a_\ell}\rb$ but the commutators can also be nested in a different way, e.g., $\lb \lb P_{a_1}, P_{a_2}\rb,\lb P_{a_3},P_{a_3}\rb\rb$. The statement here is that it is possible to find a basis of the iterated commutator form given.}
\begin{align}
\label{eq:levelstep}
\mf{f}_{\ell+1} = \lb \mf{f}_\ell,\mf{f}_1 \rb\,.
\end{align}
In appendix~\ref{app:FLA}, we review what is known about the dimensions of the space $\mf{f}_\ell$ and how free Lie algebras can be understood as special cases of generalized Kac--Moody algebras introduced by Borcherds~\cite{Borcherds}. We note also that the grading~\eqref{eq:fgrad} satisfies $\lb \mf{f}_\ell, \mf{f}_{\ell'} \rb \subset \mf{f}_{\ell+\ell'}$.

\subsection{Level decomposition}
\label{sec:LD}

We now want to make closer contact to the explicit generators introduced above in section~\ref{sec:MA}. To this end we again notice that the generating set $P_a$ spanning the $D$-dimensional space $\mf{f}_1$ can be viewed as  the fundamental of $\mf{gl}(D)$ and written as a Young tableau
\begin{align}
\mf{f}_1=\langle P_a \rangle \quad\longleftrightarrow\quad \yng(1)\,.
\end{align}

As a representation of $\mf{gl}(D)$, the next level $\mf{f}_2$ is the anti-symmetric product of two fundamental representations leading to\footnote{Here, we assume $D\geq 4$ for simplicity.}
\begin{align}
\mf{f}_2 \quad\longleftrightarrow\quad \yng(1,1)\,.
\end{align}
Introducing the corresponding tensorial generators $Z_{ab}$ of the free Lie algebra agrees with $\lb P_a, P_b\rb =Z_{ab}$ in~\eqref{eq:M2}. (This fixes a convenient normalization.)

Continuing to the next level of the free Lie algebra, we know that $\mf{f}_3$ has to be contained in the $\mf{gl}(D)$ tensor product of $\mf{f}_2$ with $\mf{f}_1$ according to~\eqref{eq:levelstep}. This tensor product is
\begin{align}
\label{eq:f213}
\yng(1,1) \otimes \yng(1) = \yng(1,1,1) \oplus \yng(2,1)\,.
\end{align}
The space $\mf{f}_3$ is a proper subset of the full tensor product since one has to impose the Jacobi identity for three elements $\lb \lb P_a, P_b\rb,P_c\rb$. The Jacobi identity $\lb \lb P_{[a}, P_b\rb, P_{c]}\rb=0$ is completely anti-symmetric and therefore $\mf{f}_3$ does not contain the totally anti-symmetric representation of $\mf{gl}(D)$, leading to
\begin{align}
\mf{f}_3 \quad\longleftrightarrow\quad \yng(2,1)\,.
\end{align}
Introducing a corresponding tensorial generator $Y_{ab,c}$, it is obtained from the lower ones by $\lb Z_{ab},P_c\rb=Y_{ab,c}$ in agreement with~\eqref{eq:L3MA}.

Four-fold commutators in the free Lie algebra must necessarily be contained in the tensor product of $\mf{f}_3$ with $\mf{f}_1$ according to~\eqref{eq:levelstep}
\begin{align}
\yng(2,1) \otimes \yng(1) = \yng(3,1)\oplus \yng(2,1,1) \oplus \yng(2,2)\,.
\end{align}
Again, not all these representation belong to $\mf{f}_4$ as one has to impose anti-symmetry and the Jacobi identity of the Lie bracket. This eliminates the last Young tableau from the above tensor product. This can be understood in general by considering the explicit multiplicities of the generators as discussed in appendix~\ref{app:FLA}. In the present case it can also be seen by realising that if this tableau were part of $\mf{f}_4$ it would arise also in the commutator $\lb\mf{f}_2,\mf{f}_2\rb$ which is anti-symmetric in the $Z_{ab}$ and can therefore cannot produce a Young tableau with shape \raisebox{0.3\height}{\scalebox{0.5}{$\yng(2,2)$}} in its commutator. Thus
\begin{align}
\mf{f}_4 \quad\longleftrightarrow\quad \yng(3,1)\oplus \yng (2,1,1)\,.
\end{align}
The corresponding generators are again those found in the cohomological approach, namely $S^1_{ab,c,d}$ and $S^2_{abc,d}$ with commutation relations shown in~\eqref{eq:L4MA}.

This process can be continued and we give the generators up to $\mf{f}_7$ in appendix~\ref{app:FLA}.

We would like to make on important remark here. For classifying the generators of the free Lie algebra $\mf{f}$, or equivalently the ones of the extensions the Poincar\'e algebra, the use of Young tableaux encoding irreducible representations of $\mf{gl}(D)$ is most convenient. However, since the are tensors of Lorentz algebra $\mf{so}(1,D-1)\subset \mf{gl}(D)$ one should properly consider the decomposition of the tensors into irreducibles of $\mf{so}(1,D-1)$. This corresponds to considering possible traces of the tensors. Starting from $\mf{f}_3$ such traces are possible.

We will use the notation that an irreducible tensor of $\mf{so}(1,D-1)$ is also denoted by a Young tableau giving the permutation symmetries of the indices. But in order to emphasize the fact that all possible traces have been removed from the tensor we will put a tilde over the tableau. For example, for $\mf{f}_3$ the decomposition of the $\mf{gl}(D)$ tensor $Y_{ab,c}$ into tensors of $\mf{so}(1,D-1)$ reads in tableau form
\begin{align}
\yng(2,1) \quad \longrightarrow\quad
\widetilde{\yng(2,1)}\oplus \yng(1)\,.
\end{align}
As tensors we write this relation as
\begin{align}
Y_{ab,c} = \tilde{Y}_{ab,c} + \frac{1}{D-1}\left(\eta_{bc} Y_a -  \eta_{ac} Y_b\right)
\end{align}
such that the trace of $Y_{ab,c}$ gives $Y_a$ and $\tilde{Y}_{ab,c}$ is traceless:
\begin{align}
\eta^{bc} Y_{ab,c} = Y_a\,,\quad
\eta^{bc} \tilde{Y}_{ab,c} =0\,.
\end{align}

The $\ell=4$ generators decompose similarly under the Lorentz group as
\begin{align}
\yng(3,1) &\quad \longrightarrow\quad \widetilde{\yng(3,1)}\oplus \widetilde{\yng(2)} \oplus \yng(1,1)\,,\nn\\
\yng(2,1,1) &\quad \longrightarrow\quad \widetilde{\yng(2,1,1)}\oplus \yng(1,1)\,,\nn\\
\end{align}
We can use this decomposition to make contact with the algebras $\mathcal{B}_n$ studied in~\cite{Salgado:2014qqa}. The algebra $\mathcal{B}_5$ contains everything on $\ell=2$ and the vector generator on $\ell=3$. The algebra $\mathcal{B}_6$ contains everything on $\ell=2$, the vector generator on $\ell=3$ and the second two-form on $\ell=4$ (the one coming from (3,1)). Thus the $\mathcal{B}_n$ algebras are subalgebras of $\Minfty$.

\subsection{Relation to Maxwell algebra}

As we have seen in the previous section, there is a close relationship between the generators appearing in the maximally extended Maxwell algebra $\Minfty$ and the those of the free Lie algebra $\mf{f}$. For levels $\ell=1,\ldots,4$ we have seen above that the correspondence is exact not only for the generators but also for the commutation relations. We will now prove that this continues to all levels. In other words, the central statement is that
\begin{align}
\Minfty \cong \mf{so}(1,D-1) \oplus \mf{f}\,,
\end{align}
where the sum is semi-direct and $\mf{f}$ is the free Lie algebra generated by the translation generators $P_a$. The isomorphism above is an isomorphism of Lie algebras. The essential part of the isomorphism is the `translation part' where the Lorentz algebra $\mf{so}(1,D-1)$ is frozen and we will restrict to this. We note that we could also introduce a dilatation operator $D$ on level $\ell=0$ that gives the level of the free Lie algebra generators as its eigenvalue:
\begin{align}
\lb D, \mf{f}_{\ell} \rb = \ell \mf{f}_\ell\,.
\end{align}
The Lorentz generators then have eigenvalue zero under $D$ since they are at level $\ell=0$.

The proof relies on studying the Eilenberg--Chevalley cohomology at level $\ell$ and goes by induction. One way of studying the cohomology is to construct the Lie algebra valued Maurer--Cartan one-form up to level $\ell$. We first need to introduce some notation. Let
\begin{align}
\Omega^{(\ell)} = g^{-1} dg = \sum_{k=1}^\ell \Omega_{(k)} \,,
\end{align}
be the Maurer--Cartan one-form up to level $\ell$, where
\begin{align}
\Omega_{(k)} = \Omega^{a_1\ldots a_k} Y_{a_1\ldots a_k}
\end{align}
is the one-form at level $k$ as signalled by the fact that the generator is a tensor with $k$ indices, meaning the generator is at level $k$ in $\Minfty$. Here, the notation does not fix the tableau structure of the indices on the level $k$ generator $Y_{a_1\ldots a_k}$. As examples, we have
\begin{align}
\Omega_{(1)} = \Omega^a P_a\,,\quad
\Omega_{(2)} = \frac12\Omega^{ab} Z_{ab}\,,\quad
\Omega_{(3)} = \frac12 \Omega^{ab,c} Y_{ab,c}\,.
\end{align}
The Maurer--Cartan equation says that
\begin{align}
d \Omega_{(k)} = - \sum_{m=1}^{k-1} \Omega_{(m)} \wedge \Omega_{(k-m)}\,,
\end{align}
in particular $d\Omega_{(1)} = 0$.

For the Eilenberg--Chevalley cohomology $H^2$ we have to study the non-trivial two-forms inductively by level. Before discussing the general case, we consider as an illustrative example $\ell=2$. All possible two-forms at this level are contained in $\Omega^{a_1a_2} \wedge \Omega^b$ and we need to find the closed but non-exact invariant ones. The action of the differential on an arbitrary two-forms is
\begin{align}
\label{eq:d21}
d \left(\Omega^{a_1a_2} \wedge \Omega^b \right) = - \Omega^{a_1} \wedge \Omega^{a_2} \wedge \Omega^b = - \Omega^{[a_1} \wedge \Omega^{a_2} \wedge \Omega^{b]} = d\left( \Omega^{[a_1a_2}\wedge \Omega^{b]}\right) \,,
\end{align}
where we have used the Maurer--Cartan equation $d\Omega^{a_1a_2} = - \Omega^{a_1} \wedge \Omega^{a_2}$ and have shown in the last steps that the differential of an arbitrary two-form is equal to that of the anti-symmetrised projection. This projection occurs in the decomposition of  the product $\Omega^{a_1a_2}\wedge \Omega^b$ into irreducible representations according to~\eqref{eq:f213}
\begin{align}
\Omega^{a_1a_2}\wedge \Omega^b = \underbrace{\Omega^{[a_1a_2}\wedge \Omega^{b]}}_{\raisebox{-0.3\height}{\scalebox{0.7}{\yng(1,1,1)}}} + \underbrace{\left(\Omega^{a_1a_2}\wedge \Omega^b -\Omega^{[a_1a_2}\wedge \Omega^{b]}\right)}_{\raisebox{-0.3\height}{\scalebox{0.7}{\yng(2,1)}}}\,.
\end{align}
Applying the dfferential $d$ to this equation and using~\eqref{eq:d21} we deduce that the structure with Young shape \raisebox{0.3\height}{\scalebox{0.5}{$\yng(2,1)$}} is closed.  Moreover, the shape \raisebox{0.3\height}{\scalebox{0.5}{$\yng(2,1)$}} is not exact since $\Omega^b$ is not the differential of anything \textit{invariant}. This last qualification is very important as we are interested in the cohomology of invariant forms. We note also that representation \raisebox{0.3\height}{\scalebox{0.5}{$\yng(1,1,1)$}} can be viewed as the Jacobi identity at this level.

After this example, we now return to the general case. In order to study the cohomology at level $\ell$ we need to consider all possible two-forms at exactly level $\ell+1$, that is linear combinations of (the overcomplete set)
\begin{align}
\Omega_{(1)} \wedge \Omega_{(\ell)}\,, \Omega_{(2)} \wedge \Omega_{(\ell-1)}\,, \ldots\,, \Omega_{(\ell)} \wedge \Omega_{(1)}\,,
\end{align}
and determine the ones that are closed but not exact. Let us focus on the last term to start with and see when it is closed:
\begin{align}
d \left( \Omega_{(\ell)} \wedge \Omega_{(1)}\right) =  d\Omega_{(\ell)} \wedge \Omega_{(1)} = -\left(\sum_{k=1}^{\ell-1}  \Omega_{(k)} \wedge \Omega_{(\ell-k)}\right) \wedge \Omega_{(1)}
\end{align}
This double commutator vanishes exactly when one can arrange the terms such that they correspond to the Jacobi identity in the free Lie algebra, similar to the example above. In other words, projecting the equation to the Jacobi identity representation does not change the result and therefore any term that satisfies the Jacobi identity will be closed while terms that do not satisfy the Jacobi identity cannot be closed.

None of the two-forms that satisfy the Jacobi identity can be exact (i.e., are differentials of an invariant one-form). If the closed two-form $\Omega_{(\ell)}\wedge \Omega_{(1)}+\ldots$ in question were exact, there would be an invariant one-form $\Theta$ with $d\Theta = \Omega_{(\ell)}\wedge\Omega_{(1)} +\ldots$. However, the form $\Omega_{(\ell)}$ is by induction assumption not exact and neither is $\Omega_{(1)}$, thus such a $\Theta$ cannot exist. This induction step for the cohomology of the Maxwell algebra mirrors the inductive statement~\eqref{eq:levelstep}. 

Therefore we have shown that the Eilenberg--Chevalley cohomology computed level by level generates exactly a free Lie algebra. This is maybe not surprising as the free Lie algebra is the maximal Lie algebra one can construct over a generating set $P_a$ and therefore there cannot be any additional extensions provided by the cohomology.

\subsection{Ideals and quotients}
\label{sec:quot}

Free Lie algebras admit many non-trivial quotient Lie algebras that arise from non-trivial ideals of the free Lie algebra. For example, the construction of standard semi-simple Lie algebras of Kac--Moody type can be viewed in this language and the class of ideals relevant there can be described in terms of Dynkin diagrams with the ideals being generated by the Serre relations~\cite{Kac}. Here, we will consider three other classes of ideals of $\mf{f}$.

The first family of ideals is defined as 
\begin{align}
\mf{i}_{\ell} =\bigoplus_{k>\ell} \mf{f}_k
\end{align}
and consists of all multi-commutators with more than $\ell$ basic generators. Since the grading~\eqref{eq:fgrad} respects the commutator this clearly is an ideal of $\mf{f}$, \textit{i.e.}, $\lb \mf{f},\mf{i}_\ell \rb \subset \mf{i}_\ell$. The associated quotient Lie algebra $\mf{q}_\ell$ is 
\begin{align}
\mf{q}_\ell = \mf{f} / \mf{i}_\ell = \bigoplus_{k=1}^\ell \mf{f}_k
\end{align}
with the same commutators as $\mf{f}$ except for that all terms leading to generators contained in $\mf{f}_k$ with $k>\ell$ are set to zero. From this we conclude that the finite Maxwell extensions of the Poincar\'e algebra satisfy
\begin{align}
\MA_\ell = \mf{so}(1,D-1) \oplus \mf{q}_\ell\,.
\end{align}

The second family of ideals is given for integers $r>0$ by 
\begin{align}
\mf{s}_r = \left\langle \textrm{Young tableaux in $\mf{f}$ with more than $r$ rows} \right\rangle\,.
\end{align}
This forms an ideal since the commutation relations of the free Lie algebra will never remove boxes from a Young tableau and therefore $\mf{s}_r$ is stable under the adjoint action of $\mf{f}$. We denote the associated quotient by
\begin{align}
\label{eq:DI}
\mf{d}_r = \mf{f}/\mf{s}_r\,.
\end{align}

The third ideal $\mf{u}$ corresponds to 
\begin{align}
\mf{u} = \left\langle \textrm{Young tableaux in $\mf{f}$ with more than $2$ rows or more than one box in the second row} \right\rangle\,.
\end{align}
The quotient
\begin{align}
\label{eq:unfold}
\mf{w} = \mf{f}/\mf{u}
\end{align}
then consists only of tableaux of the shape
\begin{align}
\young(a)\,,\quad \young(a,b)\,,\quad \young(a\cone\ctwo\cdots\cn,b)
\end{align}
for $n>0$. Except for the first tableau corresponding to $P_a$ these are exactly the tensors needed to unfold the Maxwell field strength~\cite{Boulanger:2015mka}.

There are many other ideals that could be considered for quotients. For example, any non-zero generator $x$ of $\mf{f}$ generates a non-trivial principal ideal. Other options include removing ideals generated by certain low-lying tableaux and the Serre relations of standard (Kac--Moody) Lie algebras are of this type.

\section{Point particle model with $\Minfty$ symmetry}
\label{sec:dyn}

In this section, we begin to consider a dynamical realisation of the symmetry algebra $\Minfty$ studied in the previous sections in the terms of relativistic massive and massless particle in an electro-magnetic background. For this we use the language of non-linear realisations and Lagrangian formulation also used in~\cite{Bonanos:2008kr,Bonanos:2008ez,Gibbons:2009me}.

\subsection{Coset element, Maurer--Cartan forms and Lagrangian}

We consider the coset $\Minfty / SO(1,3)$ and write a group element as\footnote{We note that our normalisations differ slightly from those employed in~\cite{Bonanos:2008ez}. For mixed symmetry Young tableau the choice of combinatorial coefficients is not fully canonical and our choice gives simple rational coefficients in the equations.}
\begin{align}
\label{eq:gpelem}
g = e^{x^a P_a} e^{\frac12\theta^{ab} Z_{ab}} e^{\frac12 \xi^{ab,c} Y_{ab,c}} e^{\frac14 \sigma_1^{ab,c,d} S_{ab,c,d}^1} e^{\frac14\sigma_2^{abc,d} S_{abc,d}^2}\cdots\,,
\end{align}
such that $X^M=(x^a,\theta^{ab},\xi^{ab,c},\sigma_1^{ab,c,d},\sigma_2^{abc,d},\ldots)$ are a choice of local coordinates on the coset space and we have fixed a Lorentz gauge. The basic building block of the non-linear realisation is the Maurer--Cartan one-form 
\begin{align}
\label{eq:CM4}
\Omega = g^{-1} dg &= d x^a P_a + \frac12\left(d\theta^{ab} +dx^a x^b\right)Z_{ab} 
  + \frac12\left(d\xi^{ab,c} - \theta^{ab} dx^c  + \frac13 dx^a x^b x^c \right) Y_{ab,c}\nn\\ 
&\quad+ \frac14 \left(d\sigma_1^{ab,c,d} -2\xi^{ab,c} dx^d +\frac16 dx^a x^b x^c x^d\right) S_{ab,c,d}^1 \nn\\
&\quad   + \frac14\left(d\sigma_2^{abc,d} +4\theta^{ab} d\theta^{cd} - 6\xi^{ab,d} dx^c -8 dx^a x^b \theta^{cd}   \right)S^2_{abc,d} +\ldots\,.
\end{align}
The form $\Omega$ is by construction invariant under all positive level elements of $\Minfty$ acting by global transformations from the left. If one wanted to keep the Lorentz gauge invariance unfixed one would have to include a factor $\exp(\tfrac12 r^{ab}M_{ab})$ in the group element~\eqref{eq:gpelem}. This could alternatively be accommodated by considering a covariant extension of the differential $d$ 
\begin{align}
d \quad \rightarrow\quad D = d + \mathcal{M}^{ab}M_{ab} \,,
\end{align}
where the last term indicates the action of the Maurer--Cartan one-form $\mathcal{M}^{ab}M_{ab}$ coming from the Lorentz piece in the group element~\cite{Bonanos:2008ez}. As we work in the fixed Lorentz gauge above, we will not require this in the sequel.

Note that the contraction with the generators automatically projects the coefficients in~\eqref{eq:CM4} on the correct Young tableau symmetries. Defining the coefficients in general as
\begin{align}
\Omega = \Omega^a P_a + \frac12 \Omega^{ab} Z_{ab} + \frac12 \Omega^{ab,c} Y_{ab,c} + \frac14 \Omega^{ab,c,d}_1 S^1_{ab,c,d} + \frac14\Omega^{abc,d}_2 S^2_{abc,d} + \ldots
\end{align}
one has the following expanded form of the projected coefficients
\begin{subequations}
\label{eq:C4exp}
\begin{align}
\Omega^a &= dx^a\,,\\
\Omega^{ab} &= d\theta^{ab}+\frac12 \left(dx^a x^b - dx^b x^a\right)\,,\\
\Omega^{ab,c} &= d\xi^{ab,c} -\frac13\left(2\theta^{ab} dx^c - \theta^{bc} dx^a - \theta^{ca} dx^b\right)
  + \frac16 \left(dx^a x^b x^c  - dx^b x^a x^c \right)\,,\\
\Omega^{ab,c,d}_1 &= d\sigma_1^{ab,c,d} -\frac34\xi^{ab,c} dx^d -\frac34 \xi^{ab,d} dx^c 
    +\frac14\left(\xi^{bc,d} + \xi^{bd,c}\right) dx^a - \frac14\left(\xi^{ac,d} + \xi^{ad,c}\right) dx^b\nn\\
    &\hspace{20mm} + \frac1{12}\left( dx^a x^b x^c x^d - dx^b x^a x^c x^d\right)\,,\\
\Omega^{abc,d}_2 &= d\sigma_2^{abc,d}  -6 dx^{[a} \xi^{bc],d} + 4 \theta^{[ab} d\theta^{c]d} -4 \theta^{[ab} d\theta^{cd]} -8 dx^{[a} x^b \theta^{c]d}  +8 dx^{[a} x^b \theta^{cd]}\,. 
\end{align}
\end{subequations}

Considering infinitesimal transformations generated by $\mf{f}$ on the lowest levels with rigid generator
\begin{align}
T = \epsilon^a P_a + \frac12 \epsilon^{ab} Z_{ab} + \frac12 \epsilon^{ab,c} Y_{ab,c} 
    +\frac14 \epsilon^{ab,c,d}_1 S^1_{ab,c,d} + \frac14 \epsilon^{abc,d}_2 S^2_{abc,d}
    + \ldots\,,
\end{align}
we find that the coset fields $X^M=(x^a,\theta^{ab},\xi^{ab,c},\sigma_1^{ab,c,d},\sigma_2^{abc,d},\ldots)$ transform as
\begin{subequations}
\begin{align}
\delta_T x^a &= \epsilon^a\,,\\
\delta_T \theta^{ab} &= \epsilon^{ab} - \frac12 \left(x^a \epsilon^b-x^b \epsilon^a\right)\,,\\
\delta_T \xi^{ab,c} &= \epsilon^{ab,c} +\frac13\left(2\epsilon^{ab} x^c -\epsilon^{bc} x^{a} -\epsilon^{ca} x^{b}\right) 
  + \frac13\left(\epsilon^a x^b x^c - \epsilon^b x^a x^c\right)\,,\\
\delta_T \sigma_1^{ab,c,d} &= \epsilon_1^{ab,c,d} + \frac34\epsilon^{ab,c} x^d + \frac34 \epsilon^{ab,d} x^c -\frac14\left(\epsilon^{bc,d}+\epsilon^{bd,c}\right) x^a +\frac14\left(\epsilon^{ac,d}+\epsilon^{ad,c}\right)x^b\nn\\
&\hspace{20mm}+\frac12\epsilon^{ab} x^c x^d -\frac14\left(\epsilon^{bc}x^a x^d - \epsilon^{ad} x^b x^c - \epsilon^{ac}x^b x^d +\epsilon^{bd} x^a x^c\right)\nn\\
&\hspace{20mm} +\frac14\left(\epsilon^a x^b x^c x^d - \epsilon^b x^a x^c x^d\right) \,,\\
\delta_T \sigma_2^{abc,d} &= \epsilon_2^{abc,d} + 6 x^{[a} \epsilon^{bc],d}  + 4 \theta^{[ab} \epsilon^{c]d} -4 \theta^{[ab} \epsilon^{cd]} +2 \epsilon^{[ab}x^{c]} x^d- 4 \epsilon^{[a} x^b \theta^{c]d}+4\epsilon^{[a} x^b \theta^{cd]}\,.
\end{align}
\end{subequations}
It is natural to think of these transformations as generalized translations in the extended space spanned by all the $X^M$. The Cartan--Maurer forms~\eqref{eq:C4exp} are invariant under these transformations.

Our next task will be to construct a (massive) particle model for motion on this coset. For this we consider the coordinates $X^M=(x^a,\theta^{ab},\xi^{ab,c},\sigma_1^{ab,c,d}, \sigma_2^{abc,d},\ldots)$ as functions of a world-line parameter $\tau$, \textit{i.e.}, $X^M\equiv X^M(\tau)$. Then the differential $d$ becomes a derivative with respect to the world-line parameter $\tau$ by the chain rule, for example $\Omega^a = \dot{x}^a d\tau$. An invariant Lagrangian up to $\ell=4$ can be constructed from the pull-backs of the one-forms $\Omega$ as~\cite{Bonanos:2008ez}
\begin{align}
\label{eq:L4}
L \, d\tau = m \sqrt{- \Omega^a \Omega_a } + \frac12 f_{ab} \Omega^{ab} + \frac12 f_{ab,c} \Omega^{ab,c}+\frac14 f_{ab,c,d}^1 \Omega^{ab,c,d}_1 + \frac14 f_{abc,d}^2 \Omega^{abc,d}_2+\ldots\,,
\end{align}
where indices are raised and lowered with the flat Minkowski metric. The new dynamical quantities $f_{ab}$, $f_{ab,c}$, $f_{ab,c,d}^1$ and $f_{abc,d}^2$ multiply the invariant one-forms and should be thought of as momentum like variables. They have the same symmetries as the corresponding generators of the free Lie algebra. 

The first Cartan form $\Omega^a=\dot{x}^a d\tau$ associated with the space-time coordinates $x^a$ plays a special role in the construction in that it is not set to zero by a Lagrange multiplier but provides a kinetic term for the particle's motion. 
One could also consider the case of massless particles by changing the first term in the Lagrangian (\ref{eq:L4}) to
$\frac{\dot x^a \dot{x}_a}{2e}$,
where $e$ is the einbein variable on the world-line. If we assign as dilatation $D$-eigenvalues the opposite dilatation weights compared to the coordinates, all terms in~\eqref{eq:L4} are invariant under $D$ except for the first term. In the massless case one can make the whole Lagrangian dilatation invariant by letting $\lb D,e\rb =-2e$.

\subsection{Equations of motion}

The Euler--Lagrange equations following from a Lagrangian of the type given in~\eqref{eq:L4} take a universal and simple form. To understand this we first consider an even simpler Lagrangian of the form
\begin{align}
L \, d\tau = f_A \Omega^A\,,
\end{align}
where $\Omega^A$ are the components of a Maurer--Cartan form $\Omega= g^{-1} dg = \Omega^A t_A$ expanded in a basis $t_A$ of a Lie algebra.The equations of motion of this system are then obtained by considering variations $g^{-1} \delta g = \Sigma^A t_A$ of the coordinates on the group manifold. Varying the Lagrangian above leads to (up to a total derivative)
\begin{align}
\delta L \, d\tau = \delta f_A \Omega^A + f_a \delta \Omega^A = \delta f_A \Omega^A - \left(\dot{f}_A + c_{AB}{}^C f_C \Omega^B \right)\Sigma^A\,,
\end{align}
in terms of the structure constants $\lb t_A, t_B \rb = c_{AB}{}^C t_C$ of the algebra. Here, we have used that the variation of the components of the Maurer--Cartan form is
\begin{align}
\delta \Omega^A = \dot{\Sigma}^A +c_{BC}{}^A \Omega^B \Sigma^C\,.
\end{align}
The equations of motion of such a system are therefore
\begin{align}
\Omega^A=0 \,,\quad \dot{f}_A = 0\,.
\end{align}

The Lagrangian~\eqref{eq:L4} does not have Lagrange multipliers $f_A$ for all Maurer--Cartan forms; the first component $\Omega^a$ is treated differently. As a result there will be associated contributions to the equations for all $\dot{f}_A$. The non-trivial Maurer--Cartan form at level $\ell=1$ is $\Omega^a=\dot{x}^a d\tau$ and therefore all equations for $\dot{f}_{\ldots}$ for levels $\ell>1$ will be proportional to $\dot{x}^a$ contracted into the Lagrange multiplier $f_{\ldots}$ of level $\ell+1$. 

To be more precise, our Lagrangian~\eqref{eq:L4} is 
\begin{align}
L \, d\tau = m\sqrt{-\Omega_a\Omega^a} + \sum_{\ell>1} f_\ell\, \Omega^\ell\,,
\end{align}
where we have used a schematic notation for the Lagrange multipliers and Maurer--Cartan forms at levels $\ell>1$. The variation of such a Lagrangian is then
\begin{align}
\delta L \, d\tau &= m \delta \sqrt{-\Omega^a\Omega_a} +\sum_{\ell>1} \delta f_\ell \Omega^\ell + \sum_{\ell>1} f_\ell \delta \Omega^\ell\nn\\
&= m  \ddot{x}_a \delta x^a d\tau + \sum_{\ell>1} \delta f_\ell  \Omega^\ell - \sum_{\ell>1}\dot{f}_\ell \Sigma^\ell - \sum_{\ell,m=1}^\infty c_{\ell m}{}^{\ell+m} f_{\ell+m} \Omega^m \Sigma^\ell \,,
\end{align}
where we have discarded total derivatives and employed proper time gauge. Moreover, the $\mathbb{Z}$-grading on the free Lie algebra was used to simplify the structure constants. The equations of motion therefore imply
\begin{align}
\label{eq:U2}
\Omega^{\ell} =0\,, \quad\quad 
\dot{f}_\ell d\tau= -c_{\ell\, 1}{}^{\ell+1} f_{\ell+1} \Omega^1\quad\quad
\textrm{ for $\ell>1$.}
\end{align}
Using the fact that $\Omega^\ell=0$ for $\ell>1$, the last term in the variation simplifies and we see that on-shell 
every $f_\ell$ only couples to the one on the next level ($f_{\ell+1}$) multiplied by $\Omega^1$ that corresponds to $\Omega^a=\dot{x}^a d\tau$. The equations for the first level are
\begin{align}
\label{eq:MU}
m\ddot{x}_a= f_{ab} \dot{x}^b\,,
\end{align}
where we have also used the commutator $\lb P_a, P_b\rb =Z_{ab}$ to make the structure constant explicit. Therefore the Lorentz equation~\eqref{eq:MU} is universal for our Lagrangian~\eqref{eq:L4} to all levels in the free Lie algebra. All the other equations~\eqref{eq:U2} are similarly universal. As an example we consider the equations of motion we work out the equation for $\dot{f}_2$. From the structure constant $\lb Z_{ab} , P_c \rb=Y_{ab,c}$ we deduce $\dot{f}_{ab} =- f_{ab,c} \dot{x}^c$. In section~\ref{sec:MP}, we will address the question to what extent $f_{ab}$ can be interpreted as an external electro-magnetic field in ordinary space-time.

Since the Lagrangian is constructed from the $\Minfty$-invariant Maurer--Cartan forms, our dynamical system has global $\Minfty$ symmetry.

\subsection{Equations of motion up to level $\ell=4$}

We now give the explicit form of the equations of motion up to level $\ell=4$ in proper time gauge. These can be calculated most easily using the universal forms derived above together with the explicit expressions~\eqref{eq:C4exp} for the Maurer--Cartan forms.  
\begin{subequations}
\label{eq:EL}
\begin{align}
\dot{\sigma}_1^{ab,c,d} &= \frac34\xi^{ab,c} \dot{x}^d +\frac34 \xi^{ab,d} \dot{x}^c 
    -\frac14\left(\xi^{bc,d} + \xi^{bd,c}\right) \dot{x}^a + \frac14\left(\xi^{ac,d} + \xi^{ad,c}\right) \dot{x}^b\nn\\
    &\quad - \frac1{24}\left( \dot{x}^a x^b c^c x^d - \dot{x}^b x^a x^c x^d\right)\,,\\
\dot{\sigma}_2^{abc,d} &= -\frac{27}{4} \dot{x}^{[a} \xi^{bc],d} - 3 \theta^{[ab} \dot{\theta}^{c]d} - 3 \theta^{d[a} \dot{\theta}^{bc]} +6 \theta^{[ab} x^{c]} \dot{x}^d +6 \theta^{d[a} x^b \dot{x}^{c]}
\,,\\
\dot{\xi}^{ab,c} &=\frac13\left( 2\theta^{ab}\dot{x}^c-\theta^{ca} \dot{x}^b-\theta^{bc} \dot{x}^a\right)
   -\frac16\left( \dot{x}^a x^b x^c- \dot{x}^b x^a x^c\right)\,,\\
\dot{\theta}^{ab} &= -\frac12\left(\dot{x}^a x^b - \dot{x}^b x^a\right)\,,\\
\dot{f}^1_{ab,c,d} &=0\,,\\
\dot{f}^2_{abc,d} &=0\,,\\
\label{eq:F3}
\dot{f}_{ab,c} &=-f^1_{ab,c,d} \dot{x}^d +\left(f^2_{abc,d} - 3f^2_{abd,c}\right)\dot{x}^d\,,\\
\label{eq:F2}
\dot{f}_{ab} 
   &= -f_{ab,c} \dot{x}^c\,,\\
\label{eq:MW4}
m \ddot{x}_a 
&=f_{ab} \dot{x}^b\,.
\end{align}
\end{subequations}
The first equations are simply the vanishing of the Maurer--Cartan forms~\eqref{eq:C4exp} enforced by the Lagrange multipliers $f_{\ldots}$ for levels $\ell>1$. If one calculated these equations directly from the Lagrangian~\eqref{eq:L4} with all expressions substituted from~\eqref{eq:C4exp} as Euler--Lagrange equations, one would of course arrive at the same result. However, the simple final expressions might appear surprising if one did not know about the underlying symmetry structure and the general considerations of the preceding section.

\subsection{Relation to multipoles}
\label{sec:MP}

Let us analyse in more detail equation~\eqref{eq:MW4} that looks like a standard Lorentz equation. The field $f_{ab}$ appearing on the right-hand side can be integrated explicitly using the equations~\eqref{eq:EL}. First, we note from~\eqref{eq:F3} that
\begin{align}
f_{ab,c} = -F_{ab,c,d}^1 x^d + \left( F^2_{abc,d} - 3 F^2_{abd,c}\right) x^d + F_{ab,c}\,,
\end{align}
where $F_{\ldots}$ are constants (along the world-line). When we consider any solution of the equations of motion the Maxwell symmetry of the dynamical system is spontaneously broken. Substituting this into~\eqref{eq:F2} one finds
\begin{align}
f_{ab} = \frac12 F^1_{ab,c,d} x^c x^d + \left( F^2_{abc,d}-3F^2_{abd,c}\right) \left(\theta^{cd} - \frac12 x^c x^d\right)
   -F_{ab,c} x^c + F_{ab}\,,
\end{align}
where $\dot{x}^c x^d = -\frac{d}{d\tau}\left(\theta^{cd} - \frac12x^c x^d\right)$ has been used. As already noted in~\cite{Bonanos:2008ez}, this `electro-magnetic field' depends on the new coordinate $\theta^{ab}$ and is not integrable in the space-time coordinates $x^a$. It is useful to separate it into an integrable part (which is a genuine electro-magnetic background) in ordinary configuration space and a non-integrable part:
\begin{align}
f_{ab} = f_{ab}^{\mathrm{int}} + f_{ab}^{\mathrm{non-int}}
\end{align}
with
\begin{subequations}
\begin{align}
f_{ab}^{\mathrm{int}}  &= \frac12 F_{ab,c,d}^1 x^c x^d - F_{ab,c} x^c + F_{ab}\,,\\
f_{ab}^{\mathrm{non-int}}  &= 4 F^2_{abc,d} \theta^{cd} + F^2_{abc,d} x^c x^d
\end{align}
\end{subequations}
The integrable part satisfies the Bianchi identity $\partial_{[a}^{\ } f_{bc]}^{\mathrm{int}} =0$.We note that the non-integrable part only depends on $F^2_{abc,d}$ whereas the integrable part looks like a Taylor expansion of an electro-magnetic field.  The full equation for $x_a$ can then be written as
\begin{align}
\label{eq:LEQ}
m\ddot{x}_a -4F^2_{abc,d}\dot{x}^b \theta^{cd} - F^2_{abc,d} \dot{\theta}^{bc} x^d = f_{ab}^{\mathrm{int}} \dot{x}^b\,.
\end{align}

For $F^2_{abc,d}=0$ one obtains the equation of motion for a particle moving in an electro-magnetic field that depends up to quadratic order on the coordinates. The condition $F^2_{abc,d}=0$ is satisfied automatically when working in the quotient $\mf{d}_2$ defined in~\eqref{eq:DI}.

If we consider the particular solution $f_{ab}^{\mathrm{int}}=0$ this would eliminate the coupling of the world-line to the electro-magnetic field and therefore should be interpreted as vanishing total charge of the system. We can, however, keep the non-integrable part in~\eqref{eq:LEQ} where the resulting equation then takes the form
\begin{align}
m\ddot{x}_a = 4F^2_{abc,d}\dot{x}^b \theta^{cd} -F^2_{abc,d} \dot{\theta}^{bc} x^d\,.
\end{align}
This equation has similarity with the equation for the motion of an electric dipole studied in~\cite{Anandan:1999ig}.

\subsection{Consistent truncation of the equations to quotients of $\Minfty$}

As discussed in section~\ref{sec:quot}, there are many quotients of $\Minfty$ that can be constructed. The dynamical equations automatically truncate consistently to any such quotient. In view of the non-integrable contributions to the Lorentz equation~\eqref{eq:LEQ} we now consider two such quotients.

The first is to work in the quotient $\mf{d}_2$ of equation~\eqref{eq:DI} where one only keeps Young tableaux with at most two rows. In this quotient up to level $\ell=4$, the non-integrable part proportional to $F^2_{abc,d}$ in~\eqref{eq:LEQ} drops out and one has a standard Lorentz equation with an electro-magnetic field that is at most cubic in the coordinate $x^a$. 

We shall also extend the equations of motion to $\ell=5$ in the quotient $\mf{d}_2$. The level five generators are given in appendix~\ref{app:CR}. The new relevant equations in the quotient $\mf{d}_2$ are
\begin{subequations}
\begin{align}
\dot{f}_{ab,c,d,e} &= 0\,,\\
\dot{f}_{ab,cd,e} &=0 \,,\\
\dot{f}_{ab,c,d} &= - f_{ab,c,d,e} \dot{x}^e - \left(f_{ab,ce,d}+f_{ab,de,c}\right)\dot{x}^e\,.
\end{align}
\end{subequations}
Integrating iteratively for $f_{ab}$ then gives the following
\begin{subequations}
\begin{align}
f_{ab,c,d} &= - F_{ab,c,d,e} x^e - \left(F_{ab,ce,d}+F_{ab,de,c}\right) x^e + F_{ab,c,d}\,,\\
f_{ab,c} &= \frac12 F_{ab,c,d,e} x^d x^e - \left(F_{ab,ce,d}+F_{ab,de,c}\right) \left(\theta^{de}-\frac12 x^dx^e\right)  -F_{ab,c,d} x^d + F_{ab,c}\,,\\
f_{ab} &= \frac16 F_{ab,c,d,e} x^c x^d x^e +\left(F_{ab,ce,d}+F_{ab,de,c}\right) \left( \xi^{de,c} -\frac16 x^c x^d x^e\right) +\frac12 F_{ab,c,d} x^c x^d - F_{ab,c} x^c +F_{ab}\,.
\end{align}
\end{subequations}
The correction to the $\xi$ term in the last line comes from 
\begin{align}
\left(\theta^{de}-\frac12 x^dx^e\right)\dot{x}^c + (c\leftrightarrow d) = \frac{d}{d\tau} \left( \xi^{de,c} -\frac16 x^c x^d x^e\right) + (c\leftrightarrow d) 
\end{align}
but seems to drop out when contracted with the $F$-terms. However, one is left with a non-integrable contribution to $f_{ab}$ given by
\begin{align}
f_{ab}^{\mathrm{non-int}}  &= \left(F_{ab,ce,d}+F_{ab,de,c}\right)  \xi^{de,c} \,.
\end{align}
This does not satisfy the Bianchi identity in the standard space-time and comes from the tableau \scalebox{0.8}{$\yng(3,2)$}. We therefore conclude that the quotient $\mf{d}_2$ introduces new additions to the Lorentz equation that are not just corresponding to the Taylor expansion of a standard electro-magnetic field. We will comment more on this in the conclusion.

The second quotient we consider is the one give in~\eqref{eq:unfold} where one only keeps Young tableaux with at most two rows such that the second row has at most one box. This will also remove the non-integrable contribution above and it is straight-forward to see that in this case one can extend the algebra and equations of motion to arbitrary level. The resulting integrable electro-magnetic field takes the form
\begin{align}
f_{ab} = \sum_{\ell\geq 0} (-1)^\ell F_{ab,i_1\ldots i_\ell} x^{i_1} \cdots x^{i_\ell}\,.
\end{align}
(The alternating sign is a consequence of our convention of appending indices on the right in the free algebra, cf.~\eqref{eq:levelstep}.) The Lorentz equation becomes simply
\begin{align}
m\ddot{x}_a = f_{ab} \dot{x}^b
\end{align}
and this quotient then describes the Taylor expansion of an electro-magnetic field. One might also refer to it as the unfolding of particle motion in an electro-magnetic field.

\section{Conclusions}
\label{sec:concl}

The Maxwell algebra  introduced in \cite{Galindo,Schrader:1972zd} is an extension of the Poincare algebra with an antisymmetric generator $Z_{ab}$.  A particle moving in a generic constant electro-magnetic background~\cite{Bonanos:2008ez} is a realisation of this algebra. The Maxwell algebra describes at same time the particle and the constant electro-magnetic background in which the particle moves. 

In this paper, we have introduced  a maximal infinite sequential extension of this algebra that we call $\Minfty$.  We have an infinite-dimensional free Lie algebra generated by the translation generators $P_a$ at level $\ell=1$ to which we join at level $\ell=0$ the Lorentz generators $M_{ab}$. The higher order levels corresponds to generators with a precise Young tableau structure. The relation with the Eilenberg--Chevalley cohomology was elucidated. The existence of different infinite Lie algebra ideals of $\Minfty$ allows construction of different truncations of the Maxwell algebra. One of these truncations corresponds  to the finite Maxwell algebras of~\cite{Bonanos:2008ez}. Another truncation gives the the unfolding of the Maxwell field given in \cite{Boulanger:2015mka,Vasiliev:2005zu}. It will be interesting to study other possible truncations of $\Minfty$.

As a possible realisation of $\Minfty$ we have constructed a model at low order in derivatives that tentatively describes the motion of a distribution of charged particles in an generic electro-magnetic field. The motion is characterised in terms of the center of mass coordinates and an infinite set of momenta, that are conjugate to the generalised coordinates of the coset $\Minfty/\mathrm{Lorentz}$. These equations take a universal form and are invariant under the $\Minfty$ symmetry by construction. By contrast any solution will break this symmetry spontaneously and the residual symmetry can be smaller, for example agreeing with the BCR algebra~\eqref{bcr1}
in the case of $\MA_2$.

We see many avenues of future research. Our treatment of the dynamical system was in Lagrangian form; for a proper analysis of the Killing symmetries of the system a transition to a canonical Hamiltonian form along the lines of~\cite{Gibbons:2009me} will be useful. The dynamical variables $f_{\ldots}$ for $\ell>1$ will then play the roles of momenta while the conjugate momentum $\pi_a$ to the position $x^a$ will take a more complicated form. The canonical formulation is also crucial for considering the potential quantisation of the system.

Two possible generalisations of the present work are to study either the non-relativistic Galilei (or Carroll) case~\cite{Bonanos:2008kr,Bergshoeff:2014jla} or the supersymmetric extension~\cite{Bonanos:2009wy}. In either case only finite extensions of the standard kinematical algebra are known but we anticipate an embedding of these structures in an appropriate free Lie algebra construction, possibly with quotients. It would also be interesting to study the relation of the electro-magnetic $\Minfty$ to other finite- or infinite-dimensional symmetries that involve gravity and/or higher rank gauge fields~\cite{Sezgin:1996cj,West:2001as,Damour:2002cu,Henneaux:2010ys}.

In conclusion, we consider our construction as providing a very general framework that can serve to analyse many physical different situations depending on which quotient of $\Minfty$ one considers.

\subsection*{Acknowledgements}

We would like to thank M.~Henneaux and J.~Palmkvist for useful discussions. This research was started during the 2015 Benasque program ``Gauge theory, supergravity and strings''. We are grateful to the Benasque center and to the program organisers for providing a stimulating environment ultimately leading to the results of this paper. AK gratefully acknowledges the warm hospitality of the University of Barcelona.
JG has been supported  in part by FPA2013-46570-C2-1-P, 2014-SGR-104 (Generalitat de Catalunya) and Consolider CPAN and by
 the Spanish goverment (MINECO/FEDER) under project MDM-2014-0369 of ICCUB (Unidad de Excelencia Mar\'\i a de Maeztu). 
\appendix

\section{Background on free Lie algebras}
\label{app:FLA}

In this appendix, we provide some more details on free Lie algebras $\mf{f}$ and how their $\mathbb{Z}$-graded decomposition
\begin{align}
\label{eq:FZ}
\mf{f} = \bigoplus_{\ell>0} \mf{f}_\ell
\end{align}
can be analysed in terms of $\mf{gl}(D)$ representations, where $D$ is the number of generators of $\mf{f}$ such that $\dim\mf{f}_1 = D$.

\subsection{Roots and multiplicities for free Lie algebras}

Before turning to the $\mathbb{Z}$-grading~\eqref{eq:FZ} one can first introduce a finer $\mathbb{Z}^D$-grading of $\mf{f}$ which resembles the root space decomposition of simple complex Lie algebras. Here, $\mathbb{Z}^D$ will play the role of the root lattice. For every one of the $D$ generators $P_a$ of $\mf{f}$ we introduce a simple root $\alpha_a$. Then we want to write $\mf{f}$ in a root space decomposition as
\begin{align}
\mf{f} = \bigoplus_{\alpha\in \mathbb{Z}^D} \mf{f}_\alpha\,.
\end{align}
The simple roots spaces are one-dimensional and spanned by the $P_a$:
\begin{align}
\mf{f}_{\alpha_a} = \langle P_a \rangle\,.
\end{align}
For $a\neq b$, the commutator $\lb P_a, P_b \rb$ is non-trivial and spans the root space of $\alpha_a+\alpha_b$:
\begin{align}
\mf{f}_{\alpha_a+\alpha_b}  = \langle \lb P_a, P_b \rb\rangle\quad\quad (a\neq b)\,.
\end{align}
For a general root $\alpha= \sum_a m_a \alpha_a$ with $m_a\geq 0$, the dimension of the root space $\mf{f}_\alpha$ is given by the Witt formula~\cite{Witt} for the multiplicity of the root:
\begin{align}
\label{eq:Witt}
\mult(\alpha)\equiv \dim \mf{f}_\alpha =  \sum_{ k | (m_0,\ldots,m_{D-1})} \frac{\mu(k)}{\height(\alpha)} \begin{pmatrix}
\frac1k \height(\alpha) \\ \frac{m_0}{k},\ldots, \frac{m_{D-1}}{k}
\end{pmatrix}\,.
\end{align}
Here, $k$ runs over all the common positive divisor of the $m_a$ and $\height(\alpha) = \sum_a m_a$ denotes the height of the root. The function $\mu$ is the M\"obius $\mu$-function:
\begin{align}
\mu(k) = \left\{\begin{array}{cl}
+1 & \textrm{if $k$ is the product of an even number of distinct prime factors (excluding $1$)}\\
-1 & \textrm{if $k$ is the product of an odd number of distinct prime factors (excluding $1$)}\\
0 & \textrm{if $k$ has a multiple prime factor (excluding $1$)}\\
\end{array}\right.
\end{align}
The polynomial (or multinomial) coefficient appearing in the Witt formula is
\begin{align}
\begin{pmatrix}
\sum t_a \\
t_0,\ldots, t_{D-1} 
\end{pmatrix} =\frac{\left(\sum t_a\right)! }{t_0! t_1! \cdots t_{D-1}!}\,.
\end{align}
Due to the relation to Borcherds algebras alluded to above one derive the Witt formula from the denominator of an appropriate Borcherds algebra~\cite{KleinschmidtThesis}. This Borcherds algebra consists solely of time-like (imaginary) simple roots without any relations among them; a possible Cartan matrix is $(-1)_{i,j=1}^D$. The associated Borcherds denominator is 
\begin{align}
\label{eq:den}
\prod_{\alpha\in Q_+} (1-e^\alpha)^{\mult(\alpha)} = 1- \sum_{a=0}^{D-1} e^{\alpha_a}\,.
\end{align}
The set $Q_+$ here denotes the set of all roots $\alpha= \sum_a m_a \alpha_a$ with $m_a \geq 0$.

As examples of the Witt formula, we work out the multiplicities $\mult(2\alpha_0)$ and $\mult(\alpha_0+\alpha_1)$ where the roots appearing are simple roots. For $2\alpha_0$ formula~\eqref{eq:Witt} evaluates to
\begin{align}
\mult(2\alpha_0)
= \frac{\mu(1)}{2} \begin{pmatrix} 2 \\2,0,0,\ldots,0 \end{pmatrix} 
  + \frac{\mu(2)}{2} \begin{pmatrix} 1 \\1,0,0,\ldots,0 \end{pmatrix} 
=  +\frac12 - \frac12 =0.
\end{align}
This is of course expected since $\lb P_0, P_0 \rb=0$ by the anti-symmetry of the Lie bracket.

By contrast, for $\alpha_0+\alpha_1$ we find
\begin{align}
\mult(\alpha_0+\alpha_1)
= \frac{\mu(1)}{2} \begin{pmatrix} 2 \\1,1,0,\ldots,0 \end{pmatrix} 
=  +\frac12 \times 2 =1 \,.
\end{align}
This is in agreement with the non-vanishing commutator $\lb P_0, P_1\rb$ in the free Lie algebra.

By specialising the denominator formula~\eqref{eq:den} one can obtain a generating function for the $\mathbb{Z}$-gradation of $\mf{f}$:
\begin{align}
\label{eq:denB}
\prod_{\ell>0} (1 - t^\ell)^{\mf{f}_\ell} = 1 - t {\mf{f}_1}.
\end{align}
The notation here is the same as in~\cite{Cederwall:2015oua} and means that one identifies the space $\mf{f}_\ell$ inside the $\mathbb{Z}^D$-gradation according to
\begin{align}
\mf{f}_\ell = \bigoplus_{\alpha\,:\, \height(\alpha)=\ell} \mf{f}_\alpha\,,
\end{align}
sometimes called the principal grading of $\mf{f}$. In order to make the notation of formula~\eqref{eq:denB} more transparent, we consider the first non-trivial term
\begin{align}
(1-t)^{\mf{f}_1}  &= (1-t e^{\alpha_0}) \cdots (1-t e^{\alpha_{D-1}}) \nn\\
 &= 1 - t (e^{\alpha_0} + \ldots + e^{\alpha_{D-1}}) + t^2 \sum_{i<j} e^{\alpha_i+\alpha_j} + O(t^3)\nn\\
 &= 1- t \mathfrak{f}_1 + t^2 \sum_{i<j} e^{\alpha_i+\alpha_j} + O(t^3)\,.
\end{align}
Multiplying this with $(1-t^2)^{\mf{f}_2}$ all terms in $t^2$ should cancel and therefore we conclude that $\mf{f}_2$ consists of all the commutators in the root spaces $\alpha_i+\alpha_j$ with $i<j$ and each root space has dimension one. This is of course in agreement with the fact that $\mf{f}_2=\lb \mf{f}_1,\mf{f}_2\rb$ and consists of all the anti-symmetric combinations.

\subsection{Weights and Young tableaux}

It is easy to implement the Witt formula and to obtain all roots on a fixed level $\ell$ with their multiplicities. We would like to present the space $\mf{f}_\ell$ as a set of Young tableaux of $\mf{gl}(D)$ in such a way that $\mf{f}_1\leftrightarrow\raisebox{-0.2\height}{\scalebox{0.8}{\yng(1)}}$. For this we use the correspondence of Young tableaux with (dominant) weights of $\mf{sl}(D)$. We will do this very explicitly for $D=4$ but the technique generalizes in a straight-forward way to other $D$. The difference between $\mf{sl}(D)$ and $\mf{gl}(D)$ only enters in the form of the constraint that all tableaux for $\mf{gl}(D)$ occuring in $\mf{f}_\ell$ have to have exactly $\ell$ boxes; thus we will explicitly show columns of $D$ boxes that have no meaning for $\mf{sl}(D)$ tensors but influence the tensor weight for $\mf{gl}(D)$ tensors.

In the following table, we list the four weights of the fundamental representation of $\mf{sl}(4)$ in Dynkin label notation and line them up against the four simple roots $\alpha_a$ of $\mf{f}_1$:
\begin{align}
\label{eq:rtswts}
\begin{tabular}{c||c|c|c|c}
root & $\alpha_0$ & $\alpha_1$ & $\alpha_2$ & $\alpha_3$\\\hline
weight & $[1,0,0]$ & $[-1,1,0]$ & $[0,-1,1]$ & $[0,0,-1]$
\end{tabular}
\end{align}
Here, we have introduced the convention that $\alpha_0$ corresponds to the highest weight of the representation. Using~\eqref{eq:rtswts} we can easily assign $\mf{sl}(4)$ weights to any root $\alpha=\sum_a m_a \alpha_a$ by linearity. For example, for we will have for the $\mf{f}_2$ root
\begin{align}
\alpha_0 + \alpha_1 = [1,0,0] + [-1,1,0] = [0,1,0]
\end{align}
and this is the highest weight of the anti-symmetric representation, as expected from the generators $Z_{ab}$ in $\raisebox{-0.2\height}{\scalebox{0.5}{\yng(1,1)}}$. 

One algorithmic way of determining the tableaux appearing at level $\ell$ in $\mf{f}$ is then to find all roots of $\mf{f}_\ell$ and to translate them to weights of $\mf{sl}(4)$. These will be the weights of a certain (possibly reducible) representation of $\mf{sl}(4)$. To find out which one it suffices to consider the dominant weights, \textit{i.e.}, those for which all entries are positive. In the $\mf{f}_1$ example above, this is only $\alpha_0$; for $\mf{f}_2$ it is only $\alpha_0+\alpha_1$. The Witt formula~\eqref{eq:Witt} gives the multiplicities of all the dominant weights and this fixes the representation of $\mf{sl}(4)$ completely. Remembering the requirement that for $\mf{f}_\ell$ all $\mf{gl}(4)$ Young tableaux should have $\ell$ boxes, the $\mf{sl}(4)$ representations can be easily converted to $\mf{gl}(4)$ representations.

We will write these dominant characters of $\mf{sl}(4)$ as $\chr(\mf{f}_\ell)$ in the following form for example for $\mf{f}_1$ and $\mf{f}_2$
\begin{align}
\chr(\mf{f}_1) = [1,0,0]\,,\quad \chr(\mf{f}_2) = [0,1,0]\,.
\end{align}
Performing the steps of the algorithm above for $\mf{f}_3$ leads to
\begin{align}
\label{eq:DC3}
\chr(\mf{f}_3) = [1,1,0] + [0,0,1]^2\,,
\end{align}
where the power $2$ for the second dominant weight indicates that it arises with multiplicity $2$. It corresponds to the root $\alpha_0+\alpha_1+\alpha_2$ that can be checked from~\eqref{eq:Witt} to have $\mult(\alpha_0+\alpha_1+\alpha_2)=2$. The character~\eqref{eq:DC3} equals the dominant character of the irreducible $\mf{sl}(4)$ representation with Young tableau 
\begin{align}
\mf{f}_3 \quad\longleftrightarrow\quad \yng(2,1)\,.
\end{align}
The dominant characters of irreducible representations of $\mf{sl}(4)$ can for example be generated using the LiE software~\cite{LiE}.

For $\mf{f}_4$ one finds
\begin{align}
\chr(\mf{f}_4) = [2,1,0]  + [0,2,0] + [1,0,1]^3 + [0,0,0]^6\,.
\end{align}
This equals the dominant character of the reducible $\mf{sl}(4)$ representation with Young tableaux
\begin{align}
\mf{f}_4 \quad\longleftrightarrow\quad \yng(3,1) \oplus \yng(2,1,1)\,.
\end{align}
To fix the decomposition it is easiest to peel off the dominant characters of the irreducible representations by size. 

Using the same method one finds the following representations
\begin{align}
\label{eq:F5}
\mf{f}_5 \quad\longleftrightarrow\quad & \yng(4,1) \oplus \yng(3,2) \oplus \yng(3,1,1) \oplus \yng(2,2,1) \oplus \yng(2,1,1,1)\,,\\[5mm]
\mf{f}_6 \quad\longleftrightarrow\quad &\yng(5,1) \oplus \yng(4,2) \oplus \yng(3,3) \oplus 2\times\yng(4,1,1) \nn\\
&\quad \oplus 3\times\yng(3,2,1) \oplus \yng(3,1,1,1) \oplus 2\times\yng(2,2,1,1)\
\end{align}
and
\begin{align}
\mf{f}_7 \quad\longleftrightarrow\quad & \yng(6,1) \oplus 2\times \yng(5,2) \oplus2\times \yng(4,3) \oplus 2\times \yng(5,1,1)\nn\\
&\quad \oplus 5\times \yng(4,2,1) \oplus 3\times \yng(3,2,2) \oplus 3\times \yng(3,3,1) \oplus 3\times \yng(4,1,1,1)\nn\\
&\quad \oplus 5\times \yng(3,2,1,1) \oplus 2\times \yng(2,2,2,1)
\end{align}

In the last two equations, some representations occur with non-trivial (outer) multiplicity.  In~\eqref{eq:F5} we see the first instance of a completion of an $\mf{sl}(4)$ representation to one of $\mf{gl}(4)$: The last diagram has a column with four boxes that is trivial for $\mf{sl}(4)$ due to the existence of the invariant $\epsilon_{a_1a_2a_3a_4}$; for $\mf{gl}(4)$ including it or not makes a difference and in the free Lie algebra it is necessary to include it.

As discussed at the end of section~\ref{sec:LD}, one could consider the decomposition of the $\mf{gl}(4)$ tensors under $\mf{so}(1,3)$ if writing everything as Lorentz group tensors.

Using the Mathematica notebook uploaded to the preprint arXiv along with this paper, one can easily find the generators at levels $\ell>7$. This notebook also makes use of the program LieLink~\cite{LieLink}.

\subsection{Commutation relations}
\label{app:CR}

We summarize the relevant commutation relations up to $\ell=5$ in the free Lie algebra for convenience. Up to $\ell=4$ they were already presented in section~\ref{sec:MA}; we denote the new generators at $\ell=5$ listed in~\eqref{eq:F5} as
\begin{align}
T_{ab,c,d,e}^1 &\quad\longleftrightarrow\quad \yng(4,1)\,,\nn\\
T_{ab,cd,e}^2 &\quad\longleftrightarrow\quad \yng(3,2)\,,\nn\\
T_{abc,d,e}^3 &\quad\longleftrightarrow\quad \yng(3,1,1)\,,\nn\\
T_{abc,de}^4 &\quad\longleftrightarrow\quad \yng(2,2,1)\,,\nn\\
T_{abcd,e}^5 &\quad\longleftrightarrow\quad \yng(2,1,1,1)\,.
\end{align}
The defining relations of the free Lie algebra are then
\begin{align}
\lb P_a, P_b \rb &= Z_{ab}\,,\nn\\
\lb Z_{ab}, P_c \rb &= Y_{ab,c}\,,\nn\\
\lb Y_{ab,c},P_d \rb &= S^1_{ab,c,d} + 2S^2_{abd,c}- S^2_{bcd,a}- S^2_{cad,b}\,,\nn\\
\lb S_{ab,c,d}^1, P_e \rb &= T_{ab,c,d,e}^1 + T_{ab,ce,d}^2 +T_{ab,de,c}^2+ 4 T_{abe,c,d}^3 - 3T^3_{e[ab,c],d} -3T^3_{e[ab,d],c}  \,,\nn\\
\lb S_{abc,d}^2, P_e \rb &= (*) T_{abc,d,e}^3  + T_{abc,de}^4 + T_{abce,d}^5+T^5_{e[abc,d]}\,.
\end{align}
The free coefficient is fixed by Jacobi identities but not required here. All remaining commutation relations are also fixed by Jacobi identities.

\end{document}